\documentclass[]{article}
\usepackage{multicol}
\usepackage{apj}

\input epsf.sty

\newcommand{\mincir}{\raise -2.truept\hbox{\rlap{\hbox{$\sim$}}\raise5.truept
\hbox{$<$}\ }}
\newcommand{\magcir}{\raise -2.truept\hbox{\rlap{\hbox{$\sim$}}\raise5.truept
\hbox{$>$}\ }}
\newcommand{\siml}{\raise -2.truept\hbox{\rlap{\hbox{$\sim$}}\raise5.truept
\hbox{$<$}\ }}
\newcommand{\simg}{\raise -2.truept\hbox{\rlap{\hbox{$\sim$}}\raise5.truept
\hbox{$>$}\ }}

\newcommand{\be}{\begin{equation}}
\newcommand{\ee}{\end{equation}}
\newcommand{\ba}{\begin{eqnarray}}
\newcommand{\ea}{\end{eqnarray}}
\newcommand{\brr}{\begin{array}}
\newcommand{\err}{\end{array}}
\newcommand{\bc}{\begin{center}}
\newcommand{\ec}{\end{center}}

\newcommand{\hm}{\,h^{-1}{\rm Mpc}}
\newcommand{\vel}{\,{\rm km\,s^{-1}}}
\newcommand{\fl}{\,{\rm erg\,s^{-1}cm^{-2}}}
\newcommand{\lum}{\,{\rm erg\,s^{-1}}}

\begin{document}

\vspace{15mm}
\begin{center}
\uppercase{Measuring $\Omega_m$ with the {\tt ROSAT} Deep Cluster Survey}\\
\vspace*{1.5ex} {\sc S. Borgani$^1$, P. Rosati$^2$, P. Tozzi$^3$,
S.A. Stanford$^4$, P.E. Eisenhardt$^5$,\\ C. Lidman$^2$, B.
Holden$^4$, R. Della Ceca$^6$, C. Norman$^7$ \& G. Squires$^8$}\\
\vspace*{1.ex}
{\small
$^1$ INFN, Sezione di Trieste, c/o Dipartimento di Astronomia
dell'Universit\`a, via Tiepolo 11, I-34100 Trieste, Italy\\
INFN, Sezione di Perugia, c/o Dipartimento di
Fisica dell'Universit\`a, via A. Pascoli, I-06123 Perugia, Italy\\
E-mail: borgani@ts.astro.it(borgani@ts.astro.it)\\
$^2$ ESO -- European Southern Observatory, D-85748 Garching
bei M\"unchen, Germany\\
E-mail:prosati@eso.org\\
$^3$ Osservatorio Astronomico di Trieste, via Tiepolo 11, I-34131
Trieste, Italy\\
E-mail: tozzi@ts.astro.it\\
$^4$ Physics Department, University of California-Davis, Davis, CA
95616, USA\\
Institute of Geophysics and Planetary Physics, Lawrence
Livermore National Laboratory\\
E-mail: adam@igpp.ucllnl.org, bholden@beowulf.ucllnl.org\\
$^5$ Jet Propulsion Laboratory, California Institute for Technology, MS
169-327, 4800 Oak Grove, Pasadena, CA 91109, USA\\
E-mail: prme@kromos.jpl.nasa.gov\\
$^6$ Osservatorio Astronomico di Brera, via Brera 28, I-20121
Milano, Italy\\
E-mail: rdc@brera.mi.astro.it\\
$^7$ Department of Physics and Astronomy, The Johns Hopkins
University, Baltimore, MD 21218, USA\\
E-mail: norman@stsci.edu\\
$^8$ Caltech Astronomy M/S 105-24, 1200 E. California Blvd., Pasadena,
CA 91125, USA\\
E-mail: gks@phobos.caltech.edu\\
}
\end{center}

\vspace*{-6pt}

\begin{abstract}
We analyze the {\tt ROSAT} Deep Cluster Survey (RDCS) to derive
cosmological constraints from the evolution of the cluster $X$--ray
luminosity distribution. The sample contains 103 galaxy clusters out
to $z\simeq 0.85$ and flux--limit $F_{lim}=3\times 10^{-14}\fl$
(RDCS-3) in the [0.5--2.0] keV energy band, with a high--redshift
extension containing four clusters at $0.90\le z\le 1.26$ and brighter
than $F_{lim}=1\times 10^{-14}\fl$ (RDCS-1). We assume cosmological
models to be specified by the matter density parameter $\Omega_m$, the
r.m.s. fluctuation amplitude at the $8\hm$ scale $\sigma_8$, and the
shape parameter for the CDM--like power spectrum $\Gamma$. Model
predictions for the cluster mass function are converted into the
$X$--ray luminosity function in two steps. First we convert mass into
intra--cluster gas temperature by assuming hydrostatic
equilibrium. Then temperature is converted into $X$--ray luminosity by
using the most recent data on the $L_X$--$T_X$ relation for nearby and
distant clusters. These include the Chandra data for seven distant
clusters at $0.57\le z\le 1.27$. From RDCS-3 we find
$\Omega_m=0.35^{+0.13}_{-0.10}$ and $\sigma_8=0.66^{+0.06}_{-0.05}$
for a spatially flat Universe with cosmological constant, with no
significant constraint on $\Gamma$ (errors correspond to $1\sigma$
confidence levels for three fitting parameters). Even accounting for
both theoretical and observational uncertainties in the mass--$X$-ray
luminosity conversion, an Einstein--de-Sitter model is always excluded
at far more than the $3\sigma$ level. We also show that the number of
$X$--ray bright clusters in RDCS-1 at $z>0.9$ are expected from the
evolution inferred at $z<0.9$ data.
\vspace*{6pt}
\noindent
{\em Subject headings: } 
Cosmology: theory - dark matter - galaxies: clusters:
general - X-rays: galaxies

\end{abstract}

\begin{multicols}{2}
\section{Introduction}
According to the standard picture of hierarchical clustering, galaxy
clusters arise from the gravitational collapse of exceptionally high
peaks of the primordial density perturbations. Therefore, they probe
the high--density tail of the distribution of the cosmic density
field, usually assumed to be Gaussian, and their number density is
exponentially sensitive to the cosmological scenario (e.g., White,
Efstathiou \& Frenk 1983). The low--redshift cluster abundance has
been used over the last decade to measure the amplitude of density
perturbations on $\sim 10\hm$ scales (here $h$ is the Hubble constant
in units of 100 $\vel$ Mpc$^{-1}$), while the redshift evolution of
the cluster abundance reflects the growth rate of density
perturbations, i.e., primarily depends on the matter density parameter
$\Omega_m$ (e.g., Oukbir \& Blanchard 1992; Eke et al. 1998).
Although very powerful in principle, this cosmological test faces two
main problems in practical applications. First, theoretical
predictions always provide the number density of clusters of a given
mass, while the mass itself is never the directly observed
quantity. Therefore, suitable assumptions are required to relate an
observational quantity to the actual cluster mass. Secondly, a cluster
sample is needed which spans a large $z$--baseline, and is based on
model-independent selection criteria, so that the search volume and
the number density associated with each cluster are uniquely
specified. In this respect, $X$--ray observations provide a very
efficient method to identify distant clusters down to a given $X$--ray
flux limit, and hence within a known survey volume for each
luminosity, $L_X$. For this reason, most of the studies which have
used clusters as cosmological probes in the literature so far are
based on $X$--ray selected samples.  Following the pioneering work
based on the Einstein Extended Medium Sensitivity Survey (EMSS, Gioia
et al. 1990; Henry et al. 1992), deep imaging data from the {\tt
ROSAT} archive have been the basis for several serendipitous,
flux-limited searches for high--redshift clusters (RDCS by Rosati et
al. 1995, 1998; 160 deg$^2$ Survey by Vikhlinin et al. 1998; SHARC by
Romer et al. 2000; WARPS by Jones et al. 1998; NEP by Gioia et
al. 2001; see also Rosati et al. 2000 and Gioia 2000 for recent
reviews).

Estimates of the $X$--ray temperature, $T_X$, for subsets of these
samples have opened the way to measure cluster masses to a fairly
good, $\sim 15\%$, precision (e.g., Evrard, Metzler \& Navarro
1996). The resulting $X$--ray temperature functions (XTF) have been
presented for both nearby (e.g., Henry \& Arnaud 1991; Markevitch
1998; see Pierpaoli, Scott \& White 2001, for a recent review) and
distant clusters (e.g., Eke et al. 1998; Donahue \& Voit 1999; Henry
2000), and have been compared with predictions from cosmological
models. The mild evolution of the XTF has been interpreted as a strong
indication for a low--density Universe, with $0.2\mincir \Omega_m
\mincir 0.6$.  However, uncertainties related to the limited amount of
high--$z$ data and to the lack of a homogeneous sample selection for
local and distant clusters could substantially weaken this conclusion
(Colafrancesco, Mazzotta \& Vittorio 1997; Viana \& Liddle 1999;
Blanchard et al. 2000).

An alternative method to estimate cluster masses is based on applying
the virial theorem to internal velocity dispersions, $\sigma_v$, as
traced by redshifts of member galaxies.  This method leads to a rather
precise determination of the mass function of nearby clusters (Girardi
et al. 1998), although it is observationally very time consuming.  The
only statistically well--defined sample of distant clusters where this
method has been applied is the CNOC survey by Carlberg et al. (1997),
which contains clusters selected from the EMSS.  Even in this case,
the limited size and redshift extension of CNOC did not allow to
stringent constraints to be placed on $\Omega_m$ from the distribution
of cluster masses (Borgani et al. 1999b).

It is worth noting also that the analyses realized so far on the
evolution of the XTF and of the $\sigma_v$--distribution combine
samples of nearby and distant cluster, which have different selection
criteria.  In principle, this could complicate the comparison between
low-- and high--redshift data when establishing the evolution of the
cluster population.

A further method to trace the evolution of the cluster number density
is to follow the evolution of the $X$--ray luminosity function (XLF).
The relation between the observed $L_X$ and the cluster virial
mass is affected by the thermodynamical status of the intra--cluster
medium (ICM). Recent observations (Ponman, Cannon \& Navarro 1999)
show an excess entropy in the ICM, not explained by gravitational
processes. This demonstrates that non--gravitational heating and,
possibly, radiative cooling significantly affects the $L_X$--$M$
relation. Despite such complexities of the ICM physics, the
$X$--ray luminosity has been shown to be a fairly robust
diagnostic of cluster masses (e.g., Borgani \& Guzzo 2001).
Furthermore, the most recent flux--limited cluster samples contain now
a large ($\sim 100$) number of objects, which are homogeneously
identified over a broad redshift baseline, out to $z\simeq 1.3$. This
provides a reliable way of combining data on nearby and distant
clusters and a straightforward estimate of the selection function.

Kitayama \& Suto (1997) and Mathiesen \& Evrard (1998) analyzed the
number counts from different $X$--ray flux--limited cluster surveys
and found that resulting constraints on $\Omega_m$ are rather
sensitive to the evolution of the mass--luminosity relation. Sadat,
Blanchard \& Oukbir (1998) and Reichart et al. (1999) analyzed the
EMSS and found results to be consistent with $\Omega_m=1$. In our
previous paper (Borgani et al. 1999, BRTN hereafter), we analyzed the
XLF from the {\tt ROSAT} Deep Cluster Survey (RDCS), as derived by
Rosati et al. (1998) at different redshift intervals, from a cluster
sample shallower than the one analyzed here (see Section 2.1,
below). We were able to set only moderate constraints on the density
parameter; we found $\Omega_m\simeq 0.4\pm 0.3$ (90\% confidence
level) for a non--evolving $L_X$--$T$ relation, with a
critical--density model still allowed by a moderate positive evolution
of this relation.  The weakness of these constraints was partly due to
the method of analysis, and partly to the smaller size of the analyzed
sample. More recently, a consistent result has been also found by
Evrard et al. (2001), who compared the RDCS redshift distribution to
results from the Hubble volume simulations.

In this paper, we present the results on cosmological constraints from
the analysis of the redshift and luminosity cluster distribution from
the final version of the RDCS. This analysis differs from that in BRTN
in several respects. First of all, the RDCS sample we analyze here is
substantially enlarged (see Section 2.1), containing more than 100
clusters selected to a fainter flux limit and out to a larger
redshift, $z\mincir 1.3$. Furthermore, in BRTN we fitted the XLF after
computing it within finite $L_X$ and $z$ intervals. The analysis
presented in this paper, instead, is based on a maximum--likelihood
method which does not rely on any binning of the data and, therefore,
has the advantage of exploiting all the information contained in the
cluster distribution within the whole portion of the $(L_X,z)$ plane
accessible to the RDCS. Finally, we base the luminosity--temperature
conversion on the most recent observations, which probe now the whole
redshift range sampled by RDCS clusters (e.g., Mushotzky \& Scharf
1997; Donahue et al. 1999; Della Ceca et al. 2000), with the notable
extension out to $z\simeq 1.3$ from the Chandra observations of the
ICM in the Lynx field (Stanford et al. 2001; Holden et al. 2001). We
will devote particular care to verifying whether and by how much
present uncertainties in the mass--luminosity relation weaken the
derived constraints on the matter density parameter and on the
amplitude of density perturbations at the cluster scale.

The plan of the paper is as follows. After briefly introducing the
RDCS sample used for this analysis, we discuss the theoretical mass
function and our approach to convert observed $X$--ray fluxes into
cluster masses. Finally, we describe in detail the maximum--likelihood
method that we apply to derive constraints on cosmological
parameters. In Section 3 we present the results of our analysis and we
discuss the main conclusions in Section 4. Unless otherwise stated,
unabsorbed fluxes and luminosities refer to the [0.5--2.0] keV energy
band.

\section{The RDCS analysis}
\subsection{The sample}
The RDCS sample was constructed from a serendipitous search for
extended sources in {\tt ROSAT} PSPC pointings with exposure longer
than 15 ksec. A wavelet-based algorithm was used to detect and measure
the angular extent of $X$--ray sources. Over 160 cluster candidates
were selected in 180 PSPC fields as sources with an extent exceeding
the local PSF with a 90\% confidence level, which was statistically
determined by a control sample of several thousands sources (Rosati et
al. 1995, 1998).

\includegraphics{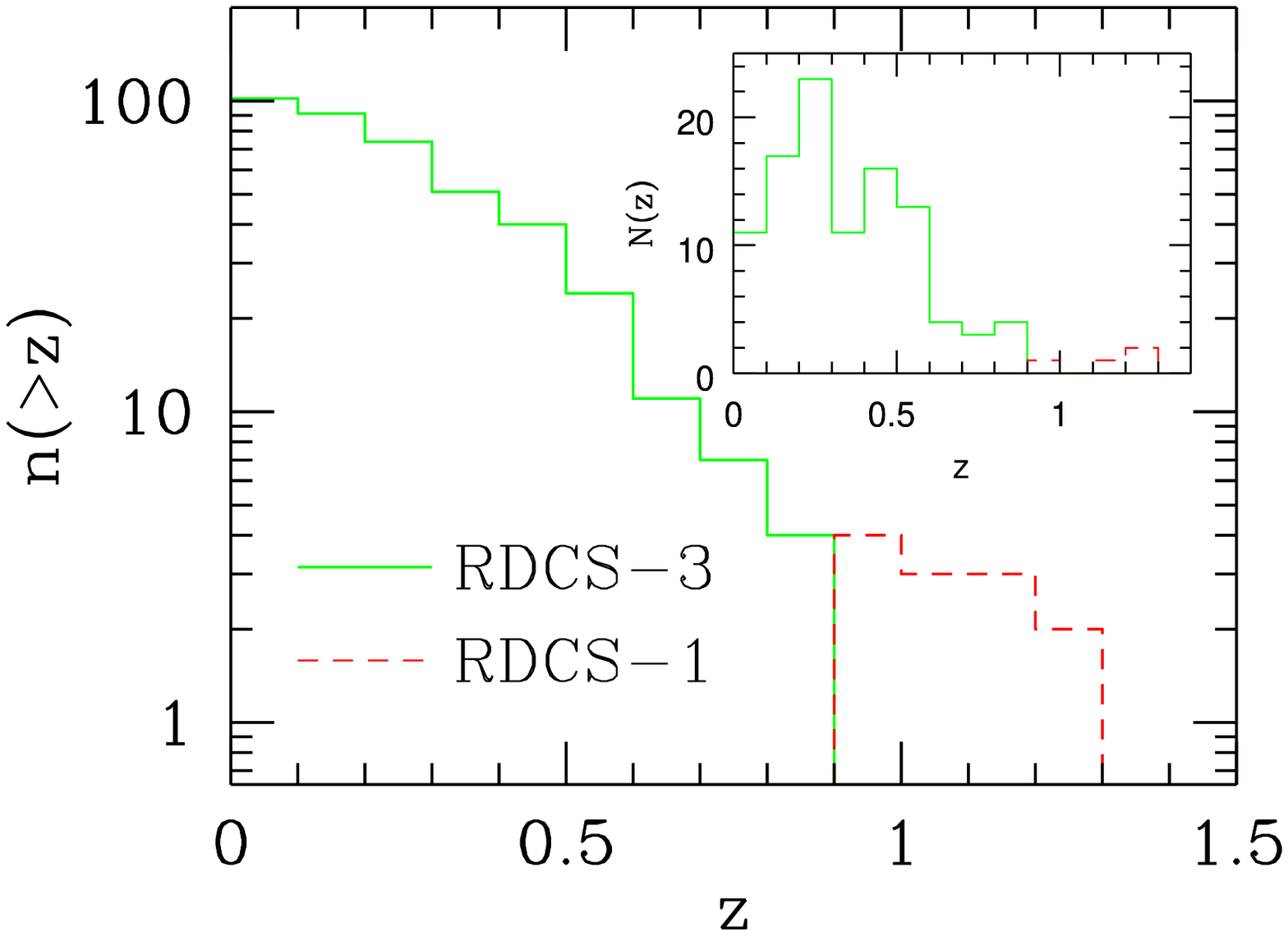} $\ \ \ \ \ \ $\\
\vspace{6.3truecm}
$\ \ \ $\\
{\small\parindent=3.5mm {Fig.}~1.---The cumulative redshift
distribution, $n(>z)$ of RDCS-3 clusters (continuous line) and of the
high--$z$ extension of the RDCS-1 clusters (dashed line). Also shown
in the insert plot is the differential redshift distribution, $N(z)$,
for the same samples.  }\vspace{5mm}

The RDCS sample that we consider in the following analysis is complete
down to the flux limit $F_{\rm lim}=3\times 10^{-14}\fl$ (RDCS-3
hereafter) and contains 103 spectroscopically confirmed clusters at
$z\le 0.85$ identified over an area of approximately 50 deg$^2$.  This
sample represents a substantial improvement with respect to that
considered by BRTN, which included 70 clusters over an area of 32
deg$^2$, above a flux-limit of $4\times 10^{-14}\fl$.  In Figure 1 we
show the cumulative and differential redshift distributions of the
RDCS-3 clusters. RDCS-3 has overall a median redshift $z_{\rm
med}=0.29$, with $z_{\rm max}=0.85$; 26 clusters lie at $z>0.5$, and 4
clusters at $z>0.8$. Several other clusters with $0.5<z<0.9$ have
been identified in the RDCS at $F_{\rm X}< 3\times 10^{-14}\fl$,
these, however, do not belong to a complete sample (see below) and
hence are not included in the present analysis. The largest flux is
$F_{\rm max}=1.24\times 10^{-12}\fl$, with a median value of the flux
distribution $F_{\rm med}=9.9\times 10^{-14}\fl$. Further details
about the sample, along with the presentation of the sky--coverage and
an analysis of the XLF evolution are presented in a separate paper
(Rosati et al., in preparation).  In addition, we also consider a
deeper subsample of four clusters identified down to $F_{lim}=1\times
10^{-14}\fl$ in the redshift range $0.90\le z\le 1.26$ (RDCS-1
hereafter), whose $z$-distribution is also shown in Fig. 1.  With the
RDCS flux-limit and sky-coverage, we find, for a critical density
Universe, that a cluster with $L_X=5\times 10^{44}\lum$ has a
searching volume $V_{\rm max}\simeq 4.5\times 10^7(\hm)^3$ at $z>0.5$
and $V_{\rm max}\simeq 3.5\times 10^7(\hm)^3$ for RDCS-3.

As discussed in Rosati et al. (1998), the sky--coverage and sample
completeness become uncertain at fluxes $\mincir 3\times
10^{-14}\fl$. This is basically due to the fact that with $\mincir 50$
counts, which roughly correspond to the above flux for the typical
exposure time of the selected PSPC fields, the detection and
characterization of extended sources becomes increasingly uncertain.
Furthermore, deep Chandra observations of the faintest RDCS clusters
in the Lynx field (Stanford et al. 2001) show that, for such faint
sources, PSPC--based $X$--ray fluxes can be significantly contaminated
by emission from foreground or background point--like sources.  For
these reasons, we base our main analysis on the relatively bright
sample, RDCS-3, while we will discuss whether the resulting
constraints are consistent with the presence of the four clusters in
the higher--$z$, lower--flux tail. Overall, our analysis will draw
information on the evolution of the cluster number density from the
widest redshift baseline presently accessible with deep cluster
surveys.

\subsection{Modeling the $L_X$ distribution}
Predictions for the $L_X$-- and $z$--distributions of RDCS clusters
are obtained by first computing the evolution of the cluster mass
function and then by converting ``theoretical'' masses into observed
$X$--ray fluxes (see also Kitayama \& Suto 1997; BRTN).

The cluster mass function is usually written as
$n(M,z)dM=(\bar\rho /M)f(\nu)(d\nu/dM)dM$, where $\bar\rho$ is the
cosmic mean density and $\nu=\delta_c/\sigma_M(z)$. Here $\delta_c$ is
the critical density contrast for top--hat halo collapse, extrapolated
at the present time by linear theory ($\delta_c=1.686$ for
$\Omega_m=1$), and $\sigma_M(z)$ the r.m.s. value of a top--hat
density fluctuation at the mass--scale $M$ at redshift $z$.  Recently,
Sheth \& Tormen (1999) have proposed for $f(\nu)$ the expression
\be f(\nu)\,=\,\sqrt{2a\over \pi}\,C\,\left(1+{1\over (a\nu^2)^q}\right)
\,\exp\left(-{a\nu^2\over 2}\right)\,,
\label{eq:st}
\ee
where $a=0.707$, $C=0.3222$ and $q=0.3$, and showed that it provides a
good fit to the halo mass function from N--body simulations (e.g.,
Governato et al. 1999; Jenkins et al. 2001). In the above equation the
normalization is determined by the requirement $\int f(\nu)d\nu =
1$. Eq.(\ref{eq:st}) reduces to the standard Press--Schechter (1974)
recipe for $a=1$, $C=1/2$ and $q=0$. We assume each cosmological model
to be specified by the matter density parameter, $\Omega_m$, and by
the CDM--like power spectrum (Bardeen et al. 1986), with profile given
by the shape--parameter $\Gamma$ and normalization by the
r.m.s. fluctuation amplitude within a sphere of $8\hm$ comoving
radius, $\sigma_8$. Unless otherwise specified, in the following we
assume flat spatial geometry provided by a cosmological constant term,
$\Omega_\Lambda=1-\Omega_m$, consistent with recent small scale
measurements of CMB anisotropies (de Bernardis et al. 2000; Hanany et
al. 2000). The value of $\Omega_\Lambda$ is known to have a relatively
small effect on the evolution of the mass function, and is far from
being constrained by current cluster samples.

Comparing a theoretical mass function with the observed $L_X$
distribution of the RDCS requires a suitable method to convert masses
into $X$--ray luminosities in the appropriate energy band, [0.5-2.0]
keV.  As a first step we convert mass into temperature by assuming
virialization, hydrostatic equilibrium, and isothermal gas
distribution, according to the relation
$k_BT=1.38\beta^{-1}\,M_{15}^{2/3}\,
\left[\Omega_m\Delta_{vir}(z)\right]^{1/3} \,(1+z)$ keV (e.g., Eke et
al. 1998).
%
%
Here we take 76\% of the gas to be hydrogen, $M_{15}$ is the cluster
virial mass in units of $10^{15}h^{-1}M_\odot$, $\beta$ the ratio
between the kinetic energy of dark matter and the gas thermal energy
($\beta = 1$ would be expected for a perfectly thermalized gas) and
$\Delta_{vir}(z)$ the ratio between the average density within the
virial radius and the mean cosmic density at redshift $z$
($\Delta_{vir}=18\pi^2\simeq 178$ for $\Omega_m=1$). Although the
assumptions on which the above relation is based have been recently
questioned (e.g., Voit 2000), such a simple approach has been shown to
reproduce fairly well the results from hydrodynamical cluster
simulations (e.g., Bryan \& Norman 1998, and references therein), with
$1\mincir \beta\mincir 1.5$.  We assume for reference the value
$\beta=1.15$ found by the Santa Barbara Cluster Comparison project
(Frenk et al. 1999).  We note that the $M$--$T_X$ relation can be
affected by the thermodynamics of the ICM.  For instance, in their
model for non--gravitational gas heating, Tozzi \& Norman (2001)
predict masses which can differ by about 20--30\% with respect to
those obtained from the above $M$--$T$ relation.  Possible deviations
with respect to this relation are also indicated by observational
data, when considering the mass within the internal cluster region at
overdensity $\delta>500$ (Finoguenov, Reiprich \& B\"ohringer
2001). Overall, we expect non--gravitational pre--heating to introduce
fairly moderate changes in the $M$--$T_X$ relation used in our
analysis. In order to account for them, in the following we will show
the effect of changing $\beta$ on the final model constraints.
Finally, we assume $15\%$ cluster-to-cluster scatter in converting
temperature into mass, as suggested by numerical simulations (Metzler,
Evrard \& Navarro 1996).

As for the relation between temperature and bolometric luminosity,
we take the phenomenological expression 
\be
L_{bol}\, = \, L_6\,\left({T_X\over 6 {\rm keV}}\right )^\alpha(1+z)^A
\left({d_L(z)\over d_{L,EdS}(z)}\right)^2
\,10^{44} h^{-2}\lum \,.
\label{eq:lt}
\ee
In this expression, $L_6$ is a dimensionless quantity and $d_L(z)$ the
luminosity--distance at redshift $z$ for a given cosmology, so that we
explicitly factorize the redshift dependence induced by changing the
spatial geometry of the cosmological background with respect to the
Einstein--de-Sitter (EdS) model. Several independent analyses of
nearby clusters with $T_X\magcir 1$ keV consistently show that
$L_6\simeq 3$ as rather stable results and $\alpha\simeq 2.5$--3
(e.g., White, Jones \& Forman 1997; Wu, Xue \& Fang 1999, and
references therein), with a rather small scatter, $\mincir 30\%$,
especially once cooling flow effects are taken into account (e.g.,
Markevitch 1998; Allen \& Fabian 1998; Arnaud \& Evrard 1999). For
cooler groups, $\mincir 1$ keV, the $L_{bol}$--$T_X$ relation
steepen, with a slope $\alpha\sim 5$ (e.g., Helsdon \& Ponman
2000). As for the redshift evolution, Mushotzky \& Scharf (1997) found
that data out to $z\simeq 0.4$ are consistent with no evolution for an
EdS model (i.e., $A\simeq 0$), a result which is consistent also with
more recent data on cluster temperatures out to $z\simeq 0.8$ (Donahue
et al. 1999; Della Ceca et al. 2000; Henry 2000).

\begin{figure*}
\includegraphics{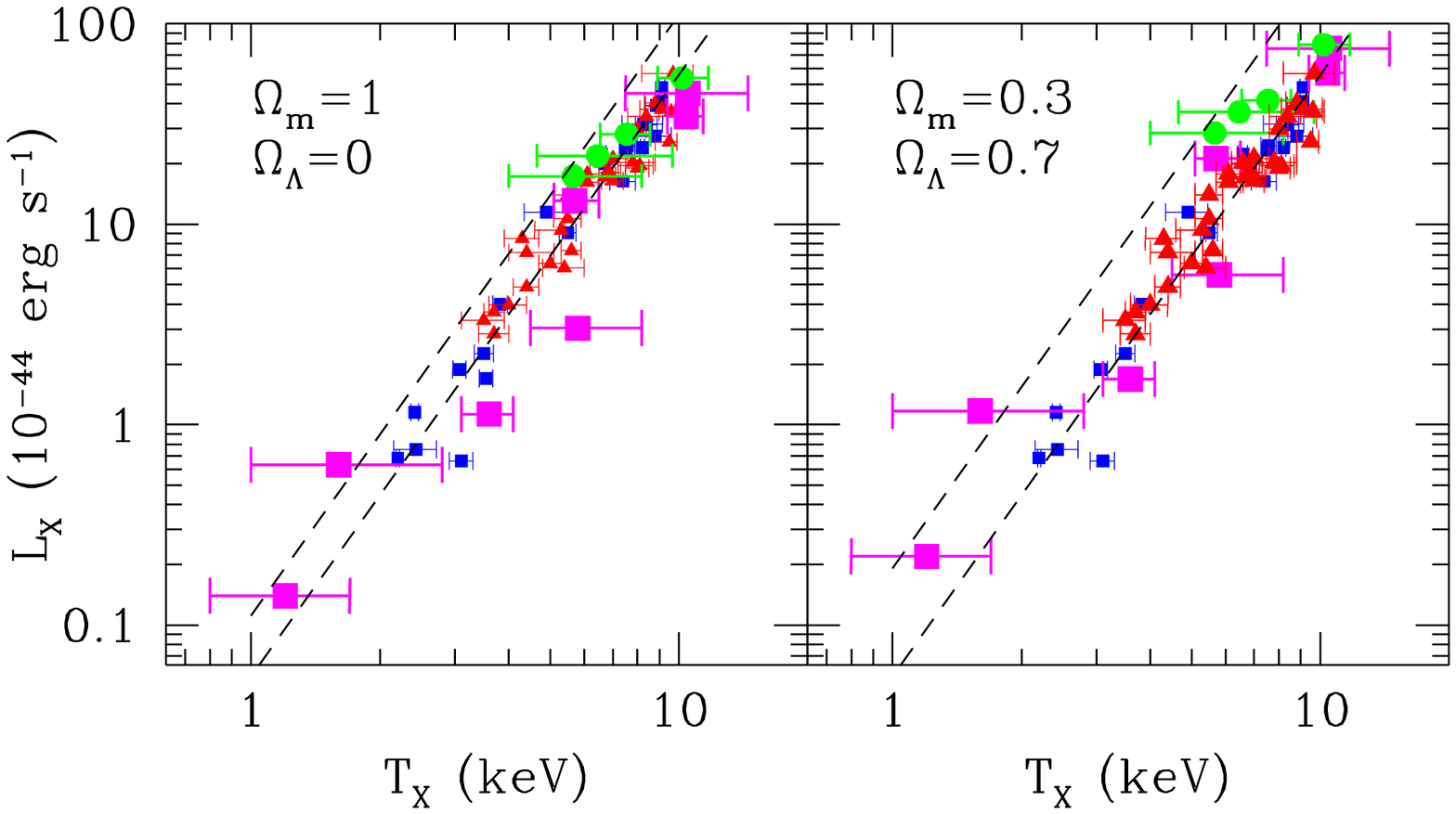}
$\ \ \ \ \ \ $\\
\vspace{6.3truecm}
$\ \ \ $\\
{\small\parindent=3.5mm {Fig.}~2.--- The luminosity--temperature relation for nearby and
distant clusters in two different cosmologies. Values of $L_X$ assume
here $h=0.5$ for the Hubble parameter. The nearby clusters analyzed by
Markevitch (1998) and by Arnaud \& Evrard (1999) are indicated with
small triangles and squares, respectively. Large circles are for the
compilation of clusters at $0.5\mincir z\mincir 0.8$ reported by Della
Ceca et al. (2000). The large squares are for the compilation of
distant ($0.57\le z\le 1.27$) clusters recently observed with Chandra
and reported in Table 1. The dashed lines indicate the
$L_{bol}$--$T_X$ relation of eq.(\ref{eq:lt}) with $L_6=3$ and
$\alpha=3$, for $A=0$ (lower lines) and $A=1$ computed at $z=1$ (upper
lines).}
\end{figure*}

%
In Table 1 we provide an update of the results available on
temperature determinations of distant ($z\ge 0.57$) clusters from
recent Chandra observations. The two $z>1$ clusters (Stanford et
al. 2001) and the $z=0.57$ cluster (Holden et al. 2001) have been
observed in the Lynx field with 190 ksec ACIS-I exposure. The cluster
CDFS-Cl1 has been detected as an extended source in the 1 Msec
observation of the Chandra Deep Field South (Giacconi et al. 2001;
Tozzi et al. 2001), its flux and temperature having been estimated
from about 350 counts (further details are presented in Giacconi et
al. in preparation). The cluster 1WGAJ1226.9 at $z=0.89$ has been
serendipitously discovered in the WARPS by Ebeling et al. (2001) and,
independently, by Cagnoni et al. (2001) in a {\tt ROSAT} PSPC blank
field. Its temperature has been measured by Cagnoni et al. with a 10
ksec ACIS-S observation. As for the cluster MS1137, it has been
observed with a 120 ksec ACIS-S pointing. We analyzed the
corresponding Chandra archival data by applying the same procedure
described by Stanford et al. (2001). The bolometric luminosity quoted
in Table 1 is computed within an aperture radius of 60''. The
resulting temperature, $T=5.7^{+0.8}_{-0.6}$ keV, turns out to be
consistent with, although somewhat more precise than, that determined
by Donahue et al. (1999) from ASCA data. Results for the cluster
MS1054 have been obtained by Jeltema et al. (2001) from a 90 ksec
exposure with Chandra ACIS-S. Also in this case, the temperature is
consistent with, although slightly smaller than that based on ASCA
observations.

The results for these distant clusters, along with other $L_X$--$T$
data for both distant and nearby clusters, are shown in Figure 2.
Overall, the results reported here constrain the evolution of the
$L_{bol}$--$T_X$ relation over the largest redshift interval probed to
date. We stress that the high redshift points always lie below or on
top of the local relation, with the possible exception of the cluster
at $z=1.27$, whose $X$--ray spatial distribution suggests this might
not be a relaxed system. In general, these data demonstrate that it is
reasonable to assume $A<1$, i.e., at most a mild positive evolution of
the $L_{bol}$--$T_X$ relation.

\vspace{6mm}
\hspace{-4mm}
\begin{minipage}{9cm}
\renewcommand{\arraystretch}{1.5}
\renewcommand{\tabcolsep}{1mm}
\begin{center}
\vspace{-3mm}
~\\ ~\\
TABLE 1\\
\footnotesize
\vspace{2mm}
\begin{tabular}{lcccc}
\hline\hline \\
Name & Redshift & \multicolumn{2}{c}{$L_X(10^{44}\lum)$} & $T$(keV) \\
     &          & EdS & $\Lambda 03$                     &        \\
\hline \\
RXJ0848+4452$^a$ & 1.26 & $3.03^{+0.83}_{-0.46}$ & $5.58^{+1.52}_{-0.85}$ & $5.8^{+2.4}_{-1.3}$  \\
RXJ0848+4453$^a$ & 1.27 & $0.63^{+0.25}_{-0.16}$ & $1.17^{+0.46}_{-0.29}$ & $1.6^{+1.2}_{-0.6}$  \\
RXJ0848+4456$^b$ & 0.57 & $1.13^{+0.34}_{-0.34}$ & $1.69^{+0.51}_{-0.51}$ & $3.6^{+0.5}_{-0.5}$  \\
CDFS-Cl1$^c$     & 0.73 & $0.14^{+0.05}_{-0.05}$ & $0.22^{+0.09}_{-0.09}$ & $1.2^{+0.5}_{-0.4}$  \\ 
1WGAJ1226.9+3332$^d$  & 0.89 & $45.0^{+4.5}_{-4.5}$   & $75.6^{+7.6}_{-7.6}$ & $10.5^{+4.0}_{-3.0}$ \\
MS1137.5+6625         & 0.78 & $13.1^{+2.6}_{-2.6}$   & $21.3^{+2.6}_{-2.6}$  & $5.7^{+0.8}_{-0.6}$  \\
MS1054.4-0321$^e$       & 0.83 & $34.5^{+3.2}_{-3.5}$ & $56.9^{+5.3}_{-5.8}$ & $10.4^{+1.0}_{-1.0}$ \\
\hline
\end{tabular}\\
\end{center}
\footnotesize{Luminosity and temperature measurements for distant ($z \ge
0.57$) clusters from recent Chandra observations. Column 1: cluster ID
($^a$ Stanford et al. 2001; $^b$ Holden et al. 2001; $^c$ Giacconi et
al. 2001, in preparation; $^d$ Redshift from Ebeling et al. 2001,
$L_X$ and $T$ from Cagnoni et al. 2001; $^e$ Jeltema et
al. 2001). Column 2: spectroscopic redshift; Columns 3-4: bolometric
luminosity, assuming $h=0.5$, for an Einstein--de-Sitter (EdS)
cosmology and for a flat low--density model with $\Omega_m=0.3$
($\Lambda 03$). Column 5: $X$--ray gas temperature and corresponding
$1\sigma$ uncertainties.}
\vspace{3mm}
\label{t:simul}
\end{minipage}

Besides the relevance for the evolution of the mass--luminosity
conversion, these results have profound implications on the physics
for the ICM. For instance, the model with constant entropy predicts
$-0.7 < A < 0.7$ depending on the level of the entropy itself and its
evolution with cosmic time (Tozzi \& Norman 2001). Values of $A$ in
this range are significantly lower than the evolution expected in the
self--similar case ($A=1.5$). Therefore, both the shape and the
non--evolution of the $L_{bol}$--$T_X$ relation are well explained in models
with substantial non--gravitational preheating.

In the following we will assume $\alpha=3$ and $A=0$ as reference
values, while we will also show the effect of changing both such
parameters.  Bolometric and $K$ corrections to the [0.5-2.0] keV
observed band are computed by using a Raymond--Smith (1977) model with
$Z=0.3$ for the mean ICM metallicity. The global scatter in converting
$X$--ray luminosity into mass is estimated by adding in quadrature a
15\% scatter in the $M$--$T_X$ and a 30\% scatter in the
$L_{bol}$--$T_X$ conversion. Its effect is then included in our
likelihood analysis by convolving the model luminosity function with a
Gaussian function having an r.m.s. scatter of 34\%. Finally, for a
given $L_X$, the flux is computed as $F=L_X/[4\pi d_L^2(z)]$.

\subsection{The analysis method}
We apply a maximum--likelihood (ML) approach, to compare the
cluster distribution in the flux--redshift, $(F,z)$, plane with
predictions from a given cosmological model for a sample having the
same flux--limit and sky--coverage as the RDCS. As a first step, we
divide the $(F,z)$ plane into elements of size $dF\,dz$ and compute
the model probability
\be
\lambda(F,z)\,dF\,dz\,=\,n[M(F),z]\,{dM\over dF}\,{dV(z)\over
dz}\,f_{sky}(F)\,dF\,dz 
\label{eq:prob}
\ee
of observing an RDCS cluster with flux $F$ at redshift $z$. Here
$dV(z)$ is the comoving volume element in the redshift interval
$[z,z+dz]$ and $f_{sky}(F)$ is the RDCS flux--dependent sky--coverage
(Rosati et al. 1998). If the bin size is small enough, such
probabilities are always much smaller than unity, then the likelihood
function ${\cal L}$ of the observed cluster flux and redshift is
defined as the product of the Poisson probabilities of observing
exactly one cluster in $dF\,dz$ at each of the $(F_i,z_i)$ positions
occupied by the RDCS clusters, and of the probabilities of observing
zero clusters elsewhere:
\be
{\cal L}\, =\, \prod_{i}\left[\lambda(F_i,z_i)\,dz\,dF\,
e^{-\lambda(F_i,z_i) dz\,dF}\right]\, \prod_{j\ne i}
e^{-\lambda(F_j,z_j)dz\,dF}\,.
\label{eq:like}
\ee
Here the indices $i$ and $j$ run over the occupied and empty elements
of the $(F,z)$ plane, respectively.  If we define the quantity
$S=-2{\mbox{\rm ln}}{\cal L}$ and drop all the terms which do not
depend on the cosmological model, it is
\be S\,=\,-2\sum_i{\mbox{\rm ln}}[\lambda(F_i,z_i)] +
2\int_0^{z_{max}} dz \int_{F_{lim}}^\infty dF\,\lambda(F,z)
\label{eq:ent}
\ee
(e.g., Marshall et al. 1983). In the above equation, $z_{max}$
represents the highest redshift at which the cluster identification
algorithm, on which RDCS is based, successfully detects extended
sources and which, in principle, does not coincide with the
highest--$z$ cluster identified in the survey. The $(1+z)^4$ surface
brightness dimming is largely responsible for this high redshift
cutoff. Using simulations as those shown in Fig.1 of Rosati et
al. (1999), we found $z_{max}=1.5$; at these redshifts the RDCS sample
becomes surface brightness limited for clusters with $L_X\simeq
10^{44}\lum$. We verified that final results are almost left
unchanged by taking $z_{max}=2$. Finally, we need to account for
non--negligible errors in cluster fluxes, which range from about 5\%
up to 35\%, with a typical value of about 15\%. To this purpose, we
take clusters not to be defined as points on the $(F,z)$ plane. Each
cluster is spread along the $F$--direction using a Gaussian smoothing
with r.m.s. amplitude equal to the flux error,
$\epsilon_F$. Therefore, instead of having zero or unity weight for
empty and occupied cells in eq.(\ref{eq:ent}), the $i$-th term
contributing to the sum is assigned the weight
\be w_{i}\,=\,\sum_{m}{1\over \sqrt{2\pi
\epsilon_{F,m}^2}}\,\exp\left[-{(F_m-F_i)^2\over
2\epsilon_{F,m}^2}\right]\,dF\,,
\label{eq:weight}
\ee
where $dF$ is its cell flux--width and the sum is over all the
clusters having redshift between $z_i-dz/2$ and $z_i+dz/2$.

\section{Results}
We derive constraints on cosmological parameters by searching for the
absolute minimum of $S$ in the three--dimensional
$(\Gamma,\sigma_8,\Omega_m)$ parameter space, and compute confidence
regions by allowing for standard increments $\Delta S$. Cosmological
parameters are varied within the following ranges: $0.02\le \Gamma \le
0.4$, $0.4\le \sigma_8\le 1.4$, $0.1\le \Omega_m\le 1$.  In Figure 3
we show the constraints on the $(\Omega_m,\sigma_8)$ plane for
different values of $\Gamma$.  As already mentioned, we assume for
this reference analysis $\beta=1.15$, $L_6=3$, $\alpha=3$ and $A=0$.
The most striking result is that the fairly large statistics and the
wide $z$--baseline provided by RDCS allow stringent constraints to be
placed, with $\Omega_m=1$ always excluded at much more than the
$3\sigma$ level.  The trend toward smaller $\Omega_m$ for larger
$\Gamma$s is due to the fact that shallower spectra produce a steeper
and more rapidly evolving mass function. Therefore, smaller $\Omega_m$
values are required to compensate for the more rapid
evolution. Although this effect is not large enough to qualitatively
affect the resulting constraints, it has nevertheless some effect on
the values of the best--fitting parameters. Overall, we obtain the
following constraints on the cosmological parameters:
\be
\Omega_m=0.35^{+0.13}_{-0.10}~~~~;~~~~\sigma_8=0.66^{+0.06}_{-0.05}
\label{eq:const}
\ee
(errors are $1\sigma$ confidence levels for three interesting
parameters). No significant constraints are found for $\Gamma$,
meaning that the sampled $L_X$ range does not correspond to a mass
range large enough to probe the shape of the power spectrum.  Indeed,
RDCS spans about 2.6 decades in $L_X$. For the adopted $M$--$L_X$
conversion, this corresponds to about 1.3 decades in mass and,
therefore, to about 0.4 decades in physical scales. For comparison,
constraints on the power--spectrum shape from the galaxy distribution
are typically derived by sampling the galaxy clustering over about two
decades in physical scales.  If we assume $\Gamma =0.25$, as suggested
by results from galaxy clustering (e.g., Dodelson \& Gazta\~naga 2000;
cf. also Efstathiou \& Moody 2000), then the resulting constraints are
$\Omega_m=0.28\pm 0.05$ and $\sigma_8=0.69\pm 0.06$, where errors are
now for two significant parameters. By assuming instead open geometry
with vanishing cosmological constant, this constraint changes into
$\Omega_m=0.35\pm 0.06$ and $\sigma_8=0.60\pm 0.03$. This shows the
tendency of flat models to favor slightly smaller values of
$\Omega_m$, as a consequence of the larger linear perturbation growth
rate in the presence of a cosmological constant term.

\begin{figure*}
\includegraphics{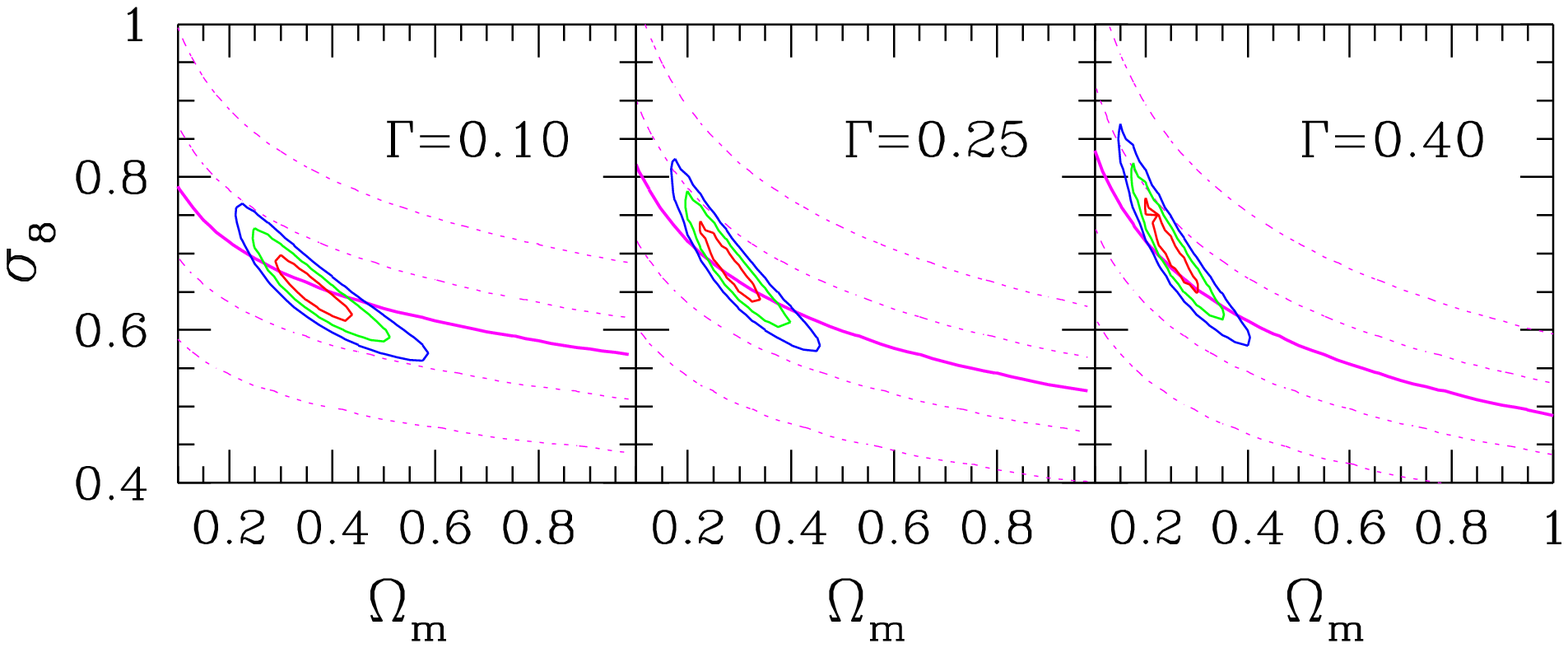}
$\ \ \ \ \ \ $\\
\vspace{5.7truecm}
$\ \ \ $\\
{\small\parindent=3.5mm {Fig.}~3.--- Confidence regions on the
$\Omega_m$--$\sigma_8$ plane for different choices of the
power--spectrum shape--parameter $\Gamma$.  Here $\alpha=3$, $A=0$ and
$\beta=1.15$ have been assumed for the mass--luminosity
conversion. Contours are $1\sigma$, $2\sigma$ and $3\sigma$ c.l. for
two interesting parameters. The dotted curves show the the
$\Omega_m$--$\sigma_8$ relations corresponding to a given number of
clusters, $N_c$, expected in RDCS-1 at $z\ge 0.9$; $N_c=0.1, 1, 10,
30$ from lower to upper curve. The solid curve correspond to the
actually observed 4 clusters.}
\end{figure*}

Also shown in Fig. 3 are the model predictions for the number of
clusters expected in RDCS-1. The different curves in each panel show
the {\em loci} of the $\Omega_m$--$\sigma_8$ plane corresponding to a
fixed number of RDCS clusters expected at $z>0.9$ and $F>1\times
10^{-14}\fl$. Quite remarkably, the four clusters detected in RDCS-1
can always be produced by cosmological models which lie inside the
$1\sigma$ contours defined by the RDCS-3. On the one hand, this result
implies that the XLF evolution traced by brighter clusters at
$z\mincir 0.85$ also extends at fainter fluxes out to the highest
redshift reached by RDCS. On the other hand, it suggests that RDCS is
not significantly affected by incompleteness or unaccounted
systematics also at fluxes $10^{-14}\fl < F_X < 3\times 10^{-14}\fl$.
It should be said, however, that given the relatively small survey
volume at these low fluxes, such systematics have in general a small
impact on observed distribution functions.

In Fig. 4 we show the effect of changing in different ways the
mass--luminosity conversion, after fixing $\Gamma=0.25$. The main
result from these tests is that $\Omega_m<0.6$ at least at the
$3\sigma$ c.l. for any variation of the mass--luminosity conversion
within realistic observational and theoretical uncertainties.  Looking
at the details of the effects, a positive evolution of the
$L_{bol}$--$T_X$ relation, which is only marginally allowed by data
(e.g., Donahue et al. 1999; Della Ceca et al. 2000; Fairley et
al. 2000; see also Fig. 2), turns into smaller masses of more distant
clusters for a fixed $L_X$. This increases the amount of evolution
inferred for the mass function and allows for a slightly larger
$\Omega_m$. A shallower slope of the local $L_{bol}$--$T_X$ relation
provides relatively smaller masses for less luminous clusters. Since
low $L_X$ objects are sampled at low redshift, this has the effect of
decreasing the amplitude of the mass function at low $z$, so as to
decrease its evolution, thus implying a lower $\Omega_m$.  We have
also verified that the observed steepening at the scale of galaxy
groups (e.g., Helsdon \& Ponman 2000) has negligible effect on our
results, only a few intermediate--$z$ RDCS clusters being faint enough
to be classified as groups. Finally, a larger $\beta$ implies a larger
cluster mass for a fixed $T_X$, so that a larger $\sigma_8$ is
required to match the amplitude of the cluster XLF. In turn, a larger
$\sigma_8$ slows down the evolution of the model mass function, thus
allowing for a larger $\Omega_m$.

These constraints on $\Omega_m$ are consistent with those
found from some analyses of the XTF evolution. 
Eke et al. (1998) combined the temperature data for 25 local clusters
by Henry \& Arnaud (1991) with the sample of 10 EMSS clusters at
$0.3<z<0.4$ by Henry (1997) and found $\Omega_m\simeq 0.40\pm 0.25$ at
the $1\sigma$ c.l. Donahue \& Voit (1999) used the low--$z$ sample by
Markevitch (1998) and enlarged the high--$z$ sample by adding further
5 clusters at $0.50\le z\le0.83$; for flat geometry, they constrained 
the density parameter to lie in the range $\Omega_m\simeq 0.3 \pm 0.1$
at the $1\sigma$ c.l. Henry (2000) combined the low--$z$ data by Henry \&
Arnaud (1991) with ASCA temperatures for 15 EMSS clusters at
$0.3<z<0.6$, and found $\Omega_m=0.44 \pm 0.12$ at the $1\sigma$
c.l. for one fitting parameter, in the case of flat geometry.

\begin{figure*}
\includegraphics{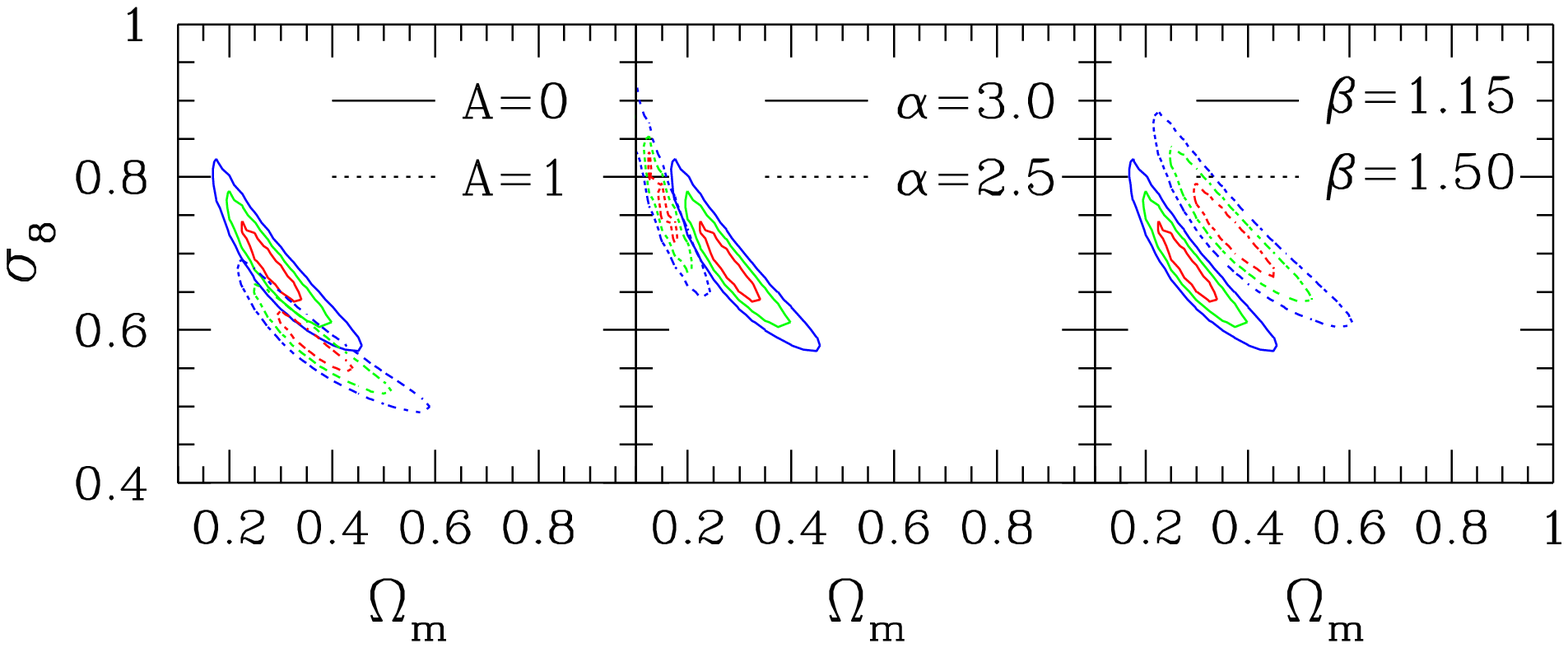}
$\ \ \ \ \ \ $\\
\vspace{5.7truecm}
$\ \ \ $\\
{\small\parindent=3.5mm {Fig.}~4.--- Effect of changing the
mass--luminosity relation.  Solid contours are from assuming
$\Gamma=0.25$, $\alpha=3$, $A=0$ and $\beta=1.15$. Each panel
corresponds to changing one of the parameters defining the
mass--luminosity relation. Contours have the same meaning as in
Fig. 3.}
\end{figure*}

As already mentioned, such constraints from the XTF could be weakened
by the lack of a homogeneous sample of cluster temperatures selected
at low and high redshift.  For instance, Viana \& Liddle (1999) used
the same data set as Eke et al. (1998) and showed that uncertainties
both in fitting local data and in the theoretical modelling could
significantly change final results. They found $\Omega_m\simeq 0.75$
as a preferred value, with a critical--density model acceptable at
$<90\%$ c.l. Blanchard et al. (2000) derived a local XTF for clusters
from the XBACs (Ebeling et al. 1996). The higher power--spectrum
normalization they derived from this sample turned into a slower XTF
evolution and, therefore, a higher $\Omega_m$. As a result, they
derived $\Omega_m\simeq 0.9 \pm 0.3$ at the $1\sigma$ c.l.  We stress
here that the possible ambiguity connected with the choice of the
low--$z$ sample and the combination of different data sets at
different redshift ranges is not an issue in our analysis, owing to
the uniform selection provided by the RDCS over its entire redshift
range.  

Besides the RDCS, the EMSS is the only other $X$-ray flux--limited sample
that has been used to date to derive cosmological constraints. Sadat
et al. (1988) used the EMSS redshift distribution, while Reichart et
al. (1999) followed the evolution of the $L_X$ distribution.  After
using a slowly evolving $L_X$--$T$ relation, they found $\Omega_m$ to
be consistent with unity and, therefore, significantly higher than the
values preferred by our RDCS analysis.

As for the amplitude of the power spectrum, our best fitting value of
$\sigma_8$ is somewhat smaller than those indicated by other
analyses. For instance, fixing $\Gamma=0.25$ and $\Omega_m=0.3$, we
find from the reference analysis shown in Fig. 3
that $\sigma_8=0.67\pm 0.06$ at the $3\sigma$ c.l. for one interesting
parameter, while Eke et al. (1998) find $\sigma_8\simeq 0.8$ with
about 20\% uncertainty. We verified that this difference can be partly
explained by the fact that we used here the mass function by Sheth
\& Tormen (1999). At the effective mass scale probed by RDCS this mass
function is somewhat larger than that by Press \& Schechter (1974),
thus requiring a lower power--spectrum normalization to match data. We
repeated our analysis with the standard Press--Schechter recipe and
found $\sigma_8=0.72$ for the same choice of $\Omega_m$
and $\Gamma$. Furthermore, we verified that, for the same
parameter choice, increasing $\beta$ to 1.25 is enough to increase
$\sigma_8$ by a further 10\%.

Finally, we point out that the constraints obtained with the present
analysis are more stringent than those we derived in BRTN.  In that
paper we used the XLF, binned in luminosity and redshift, for a
smaller version of the RDCS, and found constraints on $\Omega_m$ which
were quite dependent on the assumed evolution of the $L_X$--$T$
relation. Assuming non--evolution for this relation, we derived
$\Omega_m\simeq 0.4\pm 0.3$ at the 90\% c.l. for flat models. The much
more stringent constraints derived in the present analysis are due to
two main reasons: the larger number of clusters in the RDCS, extending
to higher redshifts and fainter fluxes; and the new analysis which
extracts more information than was previously possible with a simple
grouping of RDCS clusters into luminosity and redshift
intervals. Indeed, the likelihood function defined by
eq.(\ref{eq:ent}) conveys all the information provided by the cluster
distribution within the portion of the $(F,z)$ plane accessible to
RDCS.

\section{Conclusions}
In this paper we analyzed the evolution of the cluster number density,
as traced by the {\tt ROSAT} Deep Cluster Survey (RDCS, Rosati et
al. 1998, 2000) out to $z\simeq 1.3$, to derive constraints on the
$\Omega_m$ and $\sigma_8$ cosmological parameters. Our analysis was
aimed at understanding whether uncertainties in the relation between
cluster mass, $M$, and $X$--ray luminosity, $L_X$, prevent us from
drawing firm conclusions. In principle, a major source of uncertainty
is related to the evolution of the $M$--$T_X$--$L_X$ relation. The
most recent data shown in Fig. 2 allow us now to trace the
$L_{bol}$--$T_X$ relation out to $z\simeq 1.3$ (Stanford et al. 2001),
without evidence of significant evolution. This result also agrees
with predictions from semi--analytical models of the ICM aimed at
explaining the excess entropy observed in cluster cores (e.g., Tozzi
\& Norman 2001, and references therein). These findings are now
significantly reducing the uncertainties on the evolution of the
$X$--ray properties of cluster.

As a main result we find that, within both theoretical and
observational uncertainties in the $M$--$L_X$ relation, the density
parameter is always constrained to lie in the range $0.1\mincir
\Omega_m\mincir 0.6$ at the $3\sigma$ c.l. This demonstrates that
X-ray luminosities as a function of redshift from deep $X$--ray
flux--limited cluster surveys are indeed powerful tools to probe
cosmological scenarios.

Serendipitous cluster searches with Chandra and XMM--Newton archival
data will lead to larger distant cluster samples within the next few
years, and provide temperature information for a substantial number of
objects. Flux--limited surveys will open the way to quantify the
evolution of the cluster XLF with unprecedented accuracy, while the
availability of many more cluster temperatures, even for incomplete
samples, will provide a precise calibration of the $L_{bol}$--$T_X$
relation and its evolution.

With this perspective, the main limitation to further tightening
constraints on cosmological parameters will come from our theoretical
understanding of what a cluster actually is. While we benefited from
the recent results on the thermodynamics of the ICM for the present
analysis, another significant source of uncertainty stems from the
dynamical aspects of clusters of galaxies. For instance, the limited
validity of the assumptions on which the relation between $X$--ray
temperature and virial mass is based (e.g., spherical collapse and
hydrostatic equilibrium, see Voit 2000), makes the connection with
model predictions somewhat uncertain. Furthermore, as observations are
reaching the first epoch of cluster assembly, treating them as
dynamically relaxed and virialized systems is undoubtedly an
oversimplification. In fact, hierarchical clustering scenario predicts
that a fraction, between 0.3 and 0.6, of the $z=1$ population of
groups and clusters are observed less than 1 Gyr after the last major
merger event and, therefore, are likely to be in a state of
non--equilibrium.

Although such uncertainties have been so far of minor importance with
respect to the paucity of observational data, a breakthrough is
however needed in the quality of the theoretical framework if
high-redshift clusters are to take part in the high-precision era of
observational cosmology. In this respect, hydrodynamical cluster
simulations designed to include all the relevant ICM physics will play
a fundamental role in the years to come.

\acknowledgments We thank the anonymous referee for his/her detailed
review of the paper. SB wish to thank ESO in Garching for the hospitality
during several phases in the preparation of this work. We also thank
T. Jeltema for providing us with $L_X$ and $T$ for MS1054, from their
analysis of Chandra data, in advance of publication. Support for SAS
came from NASA/LTSA grant NAG5-8430 and for BH from NASA/Chandra
GO0-1082A, and both are supported by the Institute of Geophysics and
Planetary Physics (operated under the auspices of the US Department of
Energy by the University of California Lawrence Livermore National
Laboratory under contract W-7405-Eng-48). Portions of this work were
carried out by the Jet Propulsion Laboratory, Californian Institute of
Technology, under a contract with NASA.  Support for this work was
also provided by NASA through Hubble Fellowship Grant No.
HF-01114.01-98A from the Space Telescope Science Institute, which is
operated by the Association of Universities for Research in Astronomy,
Incorporated, under NASA Contract NAS5-26555.

\end{multicols}

\end{document}